\documentclass[graybox]{svmult}

\usepackage{amsmath}
\usepackage{type1cm}        
\usepackage{makeidx}         
\usepackage{graphicx}        
\usepackage{subcaption}                             
\usepackage{multicol}        
\usepackage[bottom]{footmisc}
\usepackage{bigstrut}
\usepackage{multirow}
\usepackage{newtxtext}       %
\usepackage[varvw]{newtxmath}       
\usepackage{float}
\usepackage{gbt7714}

\makeindex             

\begin{document}

	\title*{Nonparametric approaches for analyzing carbon emission: from statistical and machine learning perspectives} 
	\titlerunning{Nonparametric approaches for analyzing carbon emission}
	
	\author{Yiming Ma, Hang Liu and Shanyong Wang}
	\institute{Yiming Ma \at Department of Statistics and Finance, School of Management, University of Science and Technology of China  \email{mayiming@mail.ustc.edu.cn}
		\and Hang Liu  \at  International Institute of Finance, School of Management, University of Science and Technology of China \email{hliu01@ustc.edu.cn}
		\and Shanyong Wang at Department of Business Administration, School of Management, University of Science and Technology of China \email{wsy1988@ustc.edu.cn}}
	
	
	
	%
	%
	\maketitle

	\abstract{Linear regression models, especially the extended STIRPAT model, are routinely-applied for analyzing carbon emissions data. However, since the relationship between carbon emissions and the influencing factors is complex, fitting a simple parametric model may not be an ideal solution. This paper investigated various nonparametric approaches in statistics and machine learning (ML) for modeling carbon emissions data,  including kernel regression, random forest and neural network. We selected data from ten Chinese cities from 2005 to 2019 for modeling studies. We found that neural network had the best performance in both fitting and prediction accuracy, which implies its capability of expressing 
		the complex relationships between carbon emissions and the influencing factors. This study provides a new means for quantitative modeling of carbon emissions research that helps to understand how to characterize urban carbon emissions and to propose policy recommendations for ``carbon reduction''. In addition, we used the carbon emissions data of Wuhu city as an example to illustrate how to use this new approach.  }

	%

	\section{Introduction}\label{sec:intro}
	
	Climate change due to carbon emissions is causing unprecedented impacts and challenges to human society and the natural environment. Carbon emissions are greenhouse gas emissions produced in various fields and activities, mainly including carbon dioxide, methane, nitrous oxide, etc. China's carbon emissions in 2019 were about 2,777 million tons, accounting for 27\% of the world's total, making it the world's top carbon emitter \cite{larsen2021china}. On September 22, 2020, President of China Xi Jinping announced at the 75th session of the United Nations General Assembly that ``China will increase its autonomous national contribution to $\mathrm{CO}_2$  emissions will strive to peak by 2030 and work towards achieving carbon neutrality by 2060.''
	China has implemented a series of strategies, measures and actions to address climate change and participate in global climate governance.
	
	There have been numerous studies on carbon emissions  for China at the overall, regional and provincial levels \cite{chuzhi2008characteristics,du2022china}. Most of their analysis is based on the STIRPAT model and/or its extended version \cite{campbell2002proceedings,lyubich2018regulating}. STIRPAT is an important model for the study of environmental impacts, decomposing them into the products of population size, wealth per capita and technology. In this article, we consider population, affluence, energy intensity, and industrial structure as factors influencing carbon emissions.

	In previous studies, population and affluence are the two factors that most directly affect carbon emissions.  Dietz et al.\cite{dietz1997effects} believe that population and carbon emissions are proportional within a certain range, but there is a lag between policy interventions on population, and we cannot change the status of carbon emissions quickly by controlling population. In addition, they argue that carbon emissions increase and then remain constant or even decrease as the level of affluence increases. This is because being very affluent means that economic agents are transforming into a service-based economy while having enough capital to pursue energy efficiency.
	Fan et al. \cite{fan2018carbon}  study  the impact of industrial structure on carbon emissions in China and  point out that the secondary sector is the main source of carbon emissions.   
	Ang\cite{ang1999energy}  states that energy intensity is an important indicator in carbon emissions research, and  Rahman et al.\cite{rahman2022renewable} argue that reducing energy intensity and using renewable energy can effectively reduce carbon emissions. 
	
	A simple and routinely-applied approach for analyzing carbon emission data is to take a log-linear regression to  STIRPAT. However, the relationship between carbon emissions and the influencing factors is complex, and there are interactions between the influencing factors. Traditional linear regression  methods to build STIRPAT models have the disadvantage of poor fitting accuracy and inadequate explanation. For this reason, this paper aims to provide new modeling ideas and analytical tools for the study of carbon emissions. In particular, we consider various non-parametric methods, including kernel regression, which is commonly used in statistics, and random forest and neural network, which are common in ML. 
	
	We collected panel data for ten cities from 2005 to 2019 and modeled them using non-parametric methods and linear regression, finding that neural networks provide the best fit and prediction compared to linear regression method. Based on this, we have analyzed in detail the factors influencing Wuhu and made a forecast of future carbon emissions.  We found that under the premise of sacrificing a certain economic development speed, optimizing the industrial structure and improving energy efficiency, carbon emission growth can be significantly controlled.
	
	This article is organized as follows. in Section~\ref{sec:method}, we describe various methods used to model carbon emissions, including the traditional linear method and non-parametric methods (kernel regression, random forest and neural network). Section~\ref{sec:data} 
	describe the sources of the data and provide a brief description and analysis of the data. Section~\ref{sec:fitandpre} uses the four methods to model the carbon emissions data and select the best method based on the results of the fitting and prediction. Section~\ref{sec:wuhu} illustrates how to analyze the influencing factors and forecast carbon emissions based on a neural network model, using the city of Wuhu as an example. Section~\ref{sec:conclusion} concludes the paper.

	


	\section{Methodology}\label{sec:method}
	It is noted that our main objective is to predict future carbon emissions through modeling and to provide recommendations on carbon emission-related policies. We, therefore, need to model carbon emissions and the factors that influence them and expect the model to have good fit and predictive accuracy. In this section, we describe the modeling methods.

	\subsection{Extended STIRPAT model: linear regression approach} \label{sec:STIRPAT}

	IPAT model ($ I=PAT $) was first proposed by Ehrlich et al.\cite{ehrlich1971impact} and the key idea is to decompose the total environmental impact ($ I $) into the product of the population size ($ P $),  the affluence describing per capita consumption or production ({\rm A}), and the level of environmental damage caused by technology per unit of consumption or production ($ T $) \cite{lin2009analysis}. 
	On this basis,    an expandable stochastic environmental impact assessment model, called STIRPAT, was proposed by York et al.\cite{york2002bridging,york2003stirpat},
	\begin{equation}\label{STIRPAT}
		I_{it}=a_{i} P_{it}^{b} A_{it}^{c} \boldsymbol{T}_{it}^{\boldsymbol{d}} e_{it}.
	\end{equation}
	Here, $ I, P,  $ and $ A  $ are still environmental impact, population size, and affluence, respectively, but $ \boldsymbol{T}_{it} = (T_{it_1},\ldots,T_{it_q})^{\top} $ in this equation is a $q$- dimensional vector of observed variables representing technology, social organization, culture,and all other factors affecting human impact on the environment other than population and affluence, where  $\boldsymbol{A}^{\top}$ denotes the transpose of a matrix $\boldsymbol{A}$ and $q$ is a positive integer.
	$ \boldsymbol{d} = (d_1,\ldots,d_q)^{\top} $ is a $q$- dimensional vector of coefficients (or functions) representing their effects. 
	With abuse of notation, $\boldsymbol{T}_{it}^{\boldsymbol{d}}$ means $T_{it_1}^{d_1} \cdots T_{it_q}^{d_q}$.
	In addition, the subscript $ i $ and $ t $ denotes the observational units and time respectively,  and $ e $ represents sum of unexplained component and error. The coefficients $b$ and $c$ determine the net effect of population and affluence on impact, and $a$ is a constant that scales the model.
	$a,b,c,e$ and $ \boldsymbol{d} $ can be estimated by standard statistical techniques. 
	
	
	In the sequel, we will use this STIRPAT model as a benchmark model  to the study of carbon emissions.  Its performance, in terms of fitting and prediction accuracy, will be compared with our nonparametric models in Section~\ref{sec:fitandpre}.
	Let $ C $ be the carbon emission, the STIRPAT model decomposes $ C $ in the same way as equation~\eqref{STIRPAT}, that is,
	\begin{equation}\label{STIRPAT_C}
		C_{it} = a_i P^{b}_{it} A_{it}^{c} I_{it}^d E_{it}^{f} e_{it},
	\end{equation}
	where $ I $ and $ E $  denote the percentage of secondary sector and energy intensity respectively,  and $ e $ still represents the sum of unexplained components and error.

	In fact, $  I  $ represents the industrial structure, i.e. the share of the high carbon-emitting industries --- industry and construction --- in the overall economy, while $ E $ represents the level of technology (the higher the level of technology the lower the energy consumption is supposed to be). 
	
	Now we consider the problem of estimating $ a,b,c,d $ and $ f $. The standard practice is to take the logarithm of equation~\eqref{STIRPAT_C}, so the extended STIRPAT model becomes
	\begin{equation}\label{log}
		\log C_{it} = \log a + b\log P_{it}  + d \log I_{it} +f\log E_{it} + \log e_{it}.
	\end{equation}
	We can then apply the standard least squares method to estimate the coefficients. However,  there are still several problems with this estimate. (i) Time and individual differences cannot be accounted for. The fact that each coefficient is fixed means that changing the same amount of each variable has the same effect on carbon emissions. This contradicts our usual intuition. (ii) The relationship between each variable and carbon emissions may not necessarily satisfy a log-linear relationship. (iii) The variables are related to each other, which can lead to inaccurate estimates of regression coefficients, increased standard errors, wider confidence intervals and invalid significance tests.

	\subsection{Nonparametric method: from both statistical and ML perspectives}

	Rather than assuming linear or any other parametric models,  we will consider using non-parametric methods (i.e. models not specified by explicit parameters), including the use of statistical and ML methods. We note that here the non-parametric approach no longer looks for an explicit expression; we still build the regression model based on $ P,A,I $ and $ E $ suggested in model STIRPAT (i.e. input $ P,A,I $ and $ E $ and the model will output the emissions $ C $).

	\subsubsection{Non-parametric statistical method} 
	
	The kernel regression method is a non-parametric regression method used to  find a non-linear relationship between the independent and dependent variables. The essence of the kernel regression method is to use the kernel function as a weighting function to build a non-linear regression model.
	
	Let $ \boldsymbol{X}^{(n)} = \left\lbrace \boldsymbol{x}_i  \right\rbrace_{i=1}^{n}$   be a set of $d$-dimensional  vectors of observations drawn from independent variables where $ \boldsymbol{x}_i= \left(x_{i 1}, x_{i 2}, \ldots, x_{i d}\right)^{\top}$ and $d$ is a positive integer, and $y_1,y_2,\ldots,y_n$ be the corresponding response values drawn from dependent variable. Then the kernel estimator is 
	\[ 
	\hat{m}_{\boldsymbol{H}}(x) = \dfrac{\sum_{i=1}^{n} K_{\boldsymbol{H}}(\boldsymbol{x}-\boldsymbol{x}_i)y_i }{\sum_{i=1}^{n} K_{\boldsymbol{H}}(\boldsymbol{x}-\boldsymbol{x}_i)},
	\]
	where $\boldsymbol{x}=\left(x_1, x_2, \ldots, x_d\right)^{\top}$, $\boldsymbol{H} $  is the  bandwidth (or smoothing) $d \times d$-  matrix which is symmetric and positive definite, $K_{\boldsymbol{H}}$ is the kernel function  with a bandwidth $\boldsymbol{H} $.

	Briefly, the kernel regression method is a weighted average of local data points; more details can be found in  Jaakkola et al.\cite{jaakkola1999probabilistic}. In practice, we normalize the data before performing kernel regression and use the Gaussian kernel function (i.e. $K_{\boldsymbol{H}}(\boldsymbol{x})=(2 \pi)^{-d / 2}|\boldsymbol{H}|^{-1 / 2} e^{-\frac{1}{2} \boldsymbol{x}^{\top} \boldsymbol{H}^{-1} \boldsymbol{x}}$).


	\subsubsection{ML methods}
	
	\textit{Random forest}. A random forest is an integrated model consisting of multiple decision trees, which is also a weighted average of response values \cite{biau2016random}.
	Following  Athey et al. \cite{ATW19}, the weights are generated by averaging neighborhoods produced by different trees. More precisely, we grow a set of $B$ trees indexed by $b = 1, \ldots, B$ and, for each tree, let $L_b(\boldsymbol{x} )$ denote the set of training examples falling in the same leaf as $\boldsymbol{x} $. Then the weight $w_i^{(n)}(\boldsymbol{x}; \boldsymbol{X} ^{(n)}), j = 1, \ldots, n$ is the averaged (over $B$ trees) frequency that the training sample falls into the same leaf as $\boldsymbol{x}$, that is,
	$$w_i^{(n)}(\boldsymbol{x}; \boldsymbol{X}^{(n)}) = \frac{1}{B} \sum_{b=1}^B w_{bi}^{(n)}(\boldsymbol{x};  \boldsymbol{X}^{(n)}),$$
	where 
	$$w_{bi}^{(n)}(\boldsymbol{x}; \boldsymbol{X}^{(n)}) := \frac{I(\boldsymbol{X}_i^{(n)} \in L_b(\boldsymbol{x}))}{{\rm card} \{L_b(\boldsymbol{x})\}},$$
	with ${\rm card} \{L_b(\boldsymbol{x})\} $ denoting the number of elements in $L_b(\boldsymbol{x})$. Then the random forest estimator is 
	\[ 
	\hat{m}_{f}(x)=\sum_{i=1}^{n} w_i^{(n)}(\boldsymbol{x}; \boldsymbol{X}^{(n)}) y_i.
	\]
	
	\textit{Neural network}.
	A neural network is a hierarchy of interconnected nodes that are weighted and activated on the input data to obtain the final result\cite{bishop1994neural}. Specifically,
	a neural network consists of three parts: an input layer,  hidden layers and an output layer. The input layer receives the input data, the hidden layers process the input data and the output layer generates the final output. In a neural network, each node is connected to all the nodes in the next layer. Each node has a weight that is used to calculate a weighted sum of the nodes in the next layer. These weighted sums are fed into the activation function to produce the output of the nodes in the next layer. 
	
\begin{figure}[htp]
		\centering
		\includegraphics[width=\textwidth]{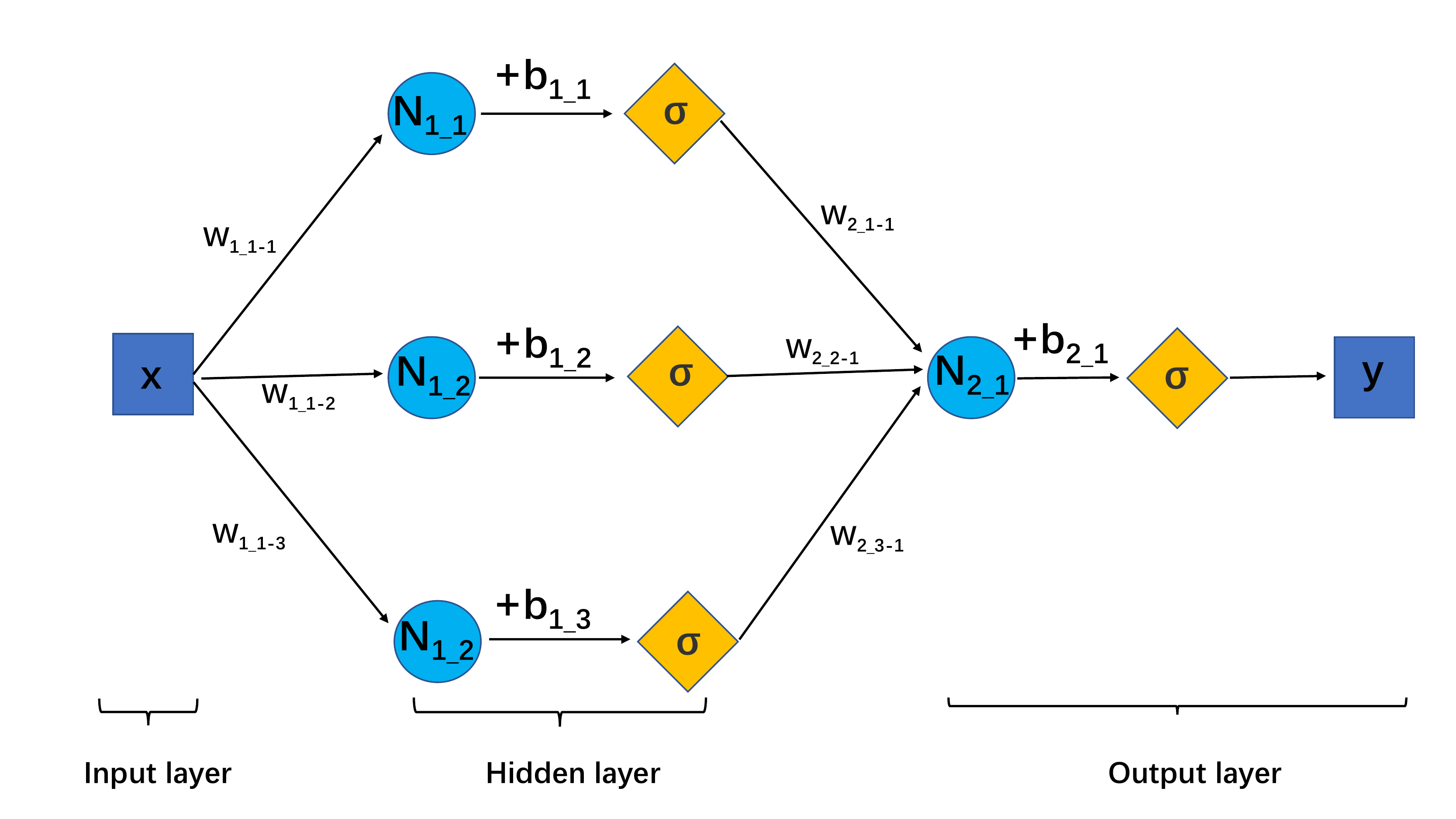}
		\caption{$w$ is the weight and $b$ is the drift. It should be noted that  the number
			in front of  sliding line ``\_" represents which layer, the number after that represents the position of the neuron, and the number after   connector ``-" represents the position of the neuron in the next layer connected. So $w_{1\_1-1}$ represents the weight between the first neuron of the first layer and the first neuron of the next layer.
			$\sigma$ is the activation function, and $ N$ is the weighted value from the neuron in the previous layer.}
		\label{fig:nn}
	\end{figure}

	We briefly introduce the workflow of  neural network through Figure~\ref{fig:nn}, which  shows a simple neural network with one hidden layer. We can observe that the input, hidden  and output layers have 1, 3 and 1 neuron(s), respectively. This neural network works as follows. We input a value $\boldsymbol{x}$, which, after linear transformations, yields the inputs of the neurons in  hidden layer.   Then, after an activation function $\sigma(\cdot)$, we obtain  three neuron outputs $\sigma(w_{1\_1-1} x + b_{1\_1}), \sigma(w_{1\_1-2} x + b_{1\_2})$ and $ \sigma(w_{1\_1-3} x + b_{1\_3})  $. The outputs of the hidden layers are  transformed linearly into the activation function $\sigma(\cdot)$ to get the final output
	$$\sigma \left[ w_{2\_1-1} \sigma(w_{1\_1-1} x + b_{1\_1} ) +  w_{2\_2-1} \sigma(w_{1\_1-2} x + b_{1\_2}) + w_{2\_3-1}  \sigma(w_{1\_1-3} x + b_{1\_3}) +b_{2\_1}\right].$$ 
	Common activation functions are the sigmod function $\operatorname{sigmoid}(x)=1/(1+e^{-x})$ and the ReLU function $\operatorname{ReLU}(x)= x I(x \geq 0)$.
	In practice, we choose the appropriate activation function according to the task. 
	If we increase the number of  hidden layers and the number of neurons per layer, then it is possible to fit complex relationships. Training a neural network is to adjust the weights and drifts so that the neural network conforms to the data distribution.

	
	Both statistical and ML nonparametric methods  can fit complex and nonlinear functional relationships. We use them in the modeling of carbon emissions.
	We expect the use of non-parametric methods to better fit the data with predicted carbon emissions so that better analysis can be carried out. In Section~\ref{sec:fitandpre}, we  compare four methods based on historical data.

	\section{Data source and description}\label{sec:data}

	\subsection{Data source}
	
	As energy consumption data at the city level is difficult to find, ten Chinese cities (including Wuhu, Hefei,  Ningbo, Guangzhou, Qingdao, Tianjin, Chongqing, Beijing, Shenzhen and Shanghai) were selected for the study. Since  CEADs \cite{guan2021assessment}  currently update city carbon emissions data to the year of 2019, we selected the data of these ten cities from 2005 to 2019. Specifically, note that $ A = \frac{\mathrm{GDP}}{P} $, $ I=\frac{\rm GDP_{ind}}{\mathrm{GDP}} $ and $ E=\frac{En}{\rm GDP} $, so we should collect the data of $ C, P,{\rm  GDP, GDP_{ind}} $ and $ En $. The description and sources of these data are listed in Table~\ref{tab:1}.

	\begin{table}[htbp]
		\centering
		\resizebox{.9\columnwidth}{!}{
			\begin{tabular}{llll}
				\hline
				Data  & Definition  & Source & Unit \bigstrut\\
				\hline
				$ C  $    & carbon emission ($ \mathrm{CO}_2 $) &  \multicolumn{1}{p{23em}}{Carbon accounts and datasets (CEADs); \newline Dataset URL: \url{https://www.ceads.net/  }  }  & ton \bigstrut\\
				\hline
				$ \mathrm{GDP}  $  & gross secondary industry & Statistical Yearbook for each municipality  & CNY \bigstrut\\
				\hline
				$ \mathrm{GDP_{ind}}  $& gross secondary industry & Statistical Yearbook for each municipality & CNY \bigstrut\\
				\hline
				$ En   $  & energy consumption & \multicolumn{1}{p{23em}}{Statistical yearbooks for each municipality (Ningbo \newline energy  data is from Ningbo Energy White Paper)}  &  ton \bigstrut\\
				\hline
				$ P  $    & Resident population & Statistical yearbooks for each municipality & Number \bigstrut\\
				\hline
			\end{tabular}%
		}
		\caption{Definition and source of the data}
		\label{tab:1}%
	\end{table}%
	
	In order not to cause misunderstanding, a few points need to be clarified: (i) GDP and $\mathrm{GDP_{ind}}$ is calculated using constant 2005 prices; (ii) energy consumption has been converted to standard coal and (iii) data is missing for some cities and years. For example, we could not find the energy consumption data of Shenzhen from 2005 to 2009 and the carbon emission data of Wuhu in 2005.
	
	
	\subsection{Overview of the data}
	
\begin{figure}[htp]
		\centering
		\begin{subfigure}[b]{0.49\textwidth}
			\centering
			\includegraphics[width=\textwidth]{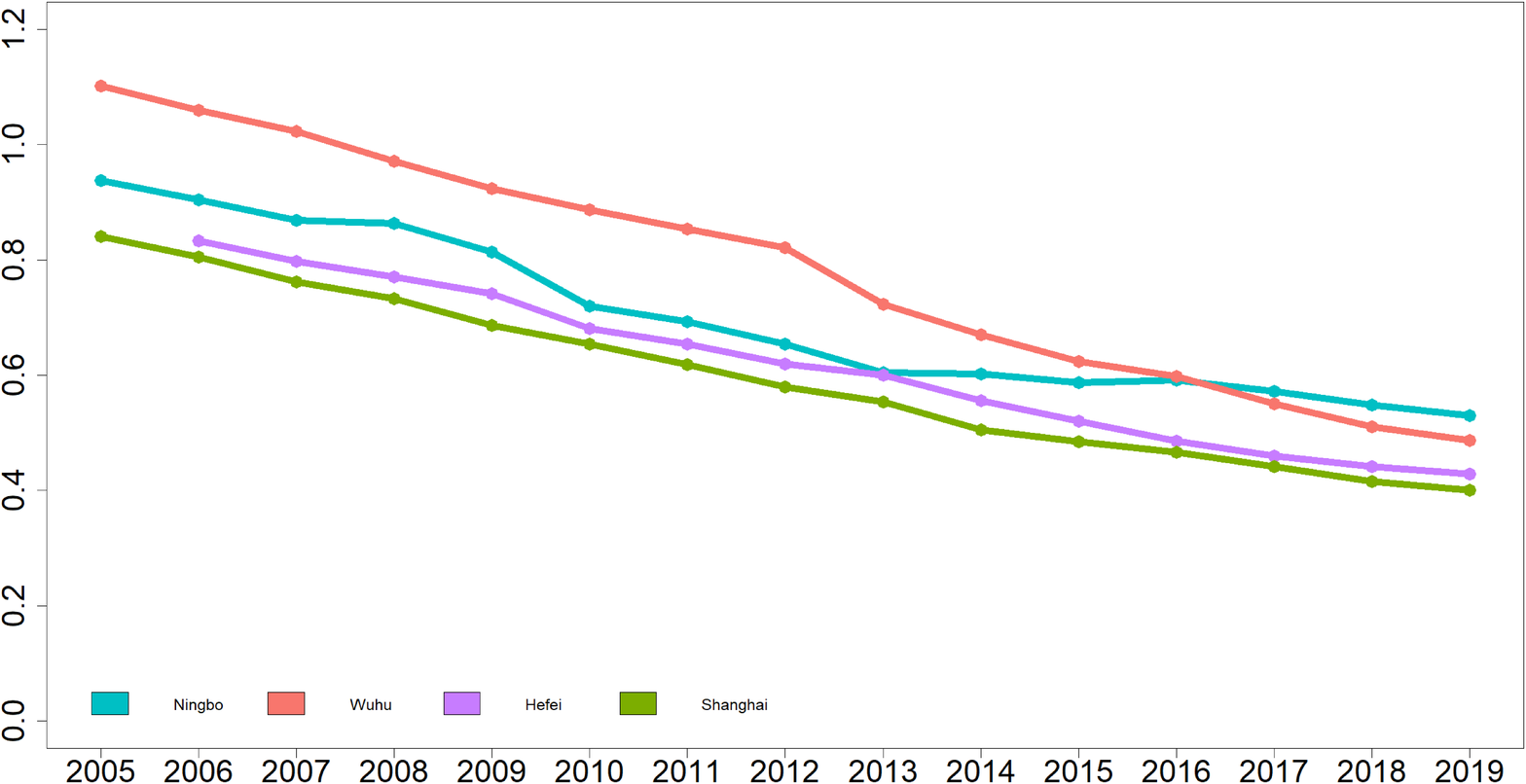}
			\caption{Energy intensity (ton/10,000 CNY) }
		\end{subfigure}
		\begin{subfigure}[b]{0.49\textwidth}
			\centering
			\includegraphics[width=\textwidth]{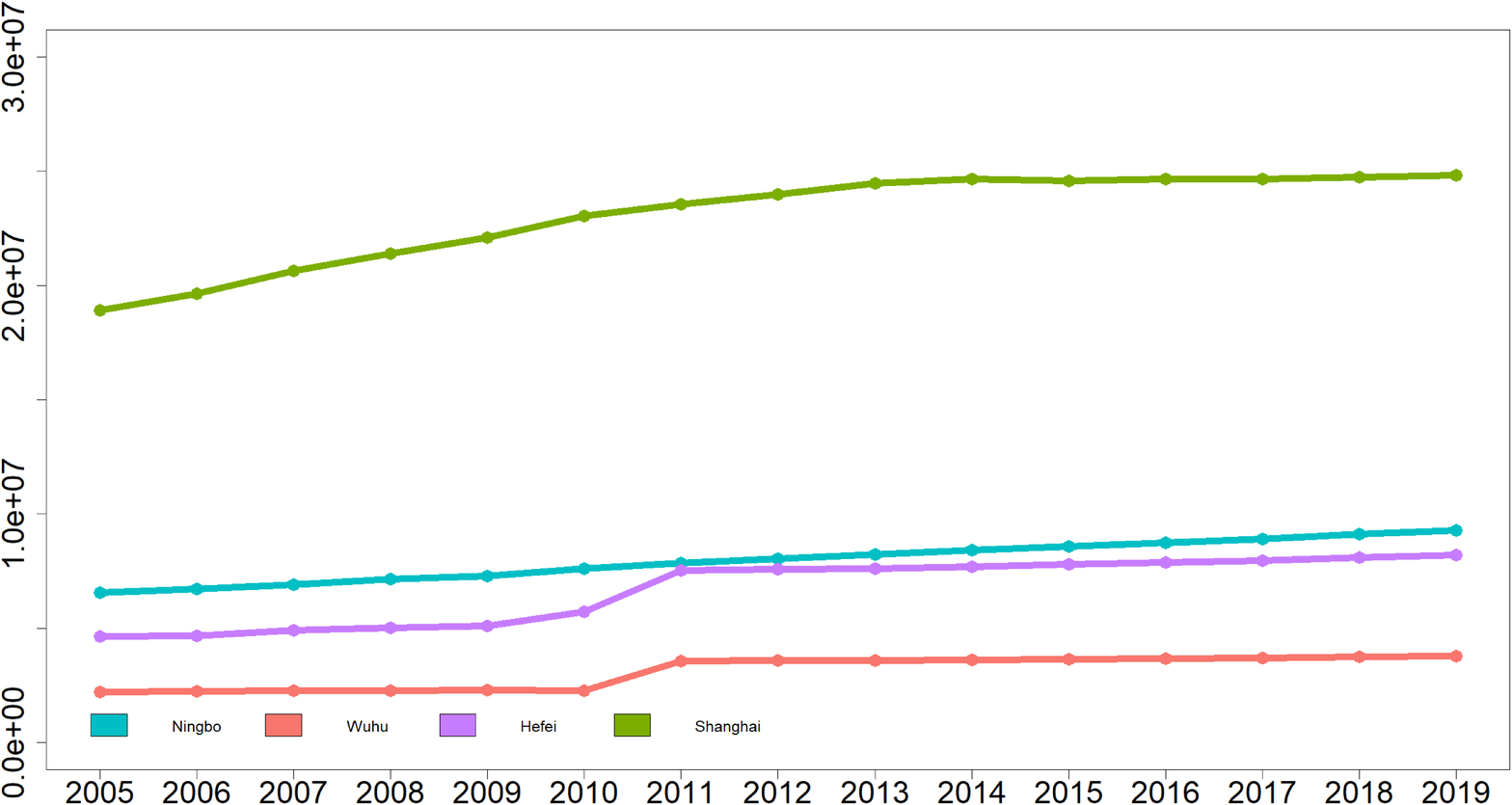}
			\caption{Population (Number)}
		\end{subfigure}
		\begin{subfigure}[b]{0.49\textwidth}
			\centering
			\includegraphics[width=\textwidth]{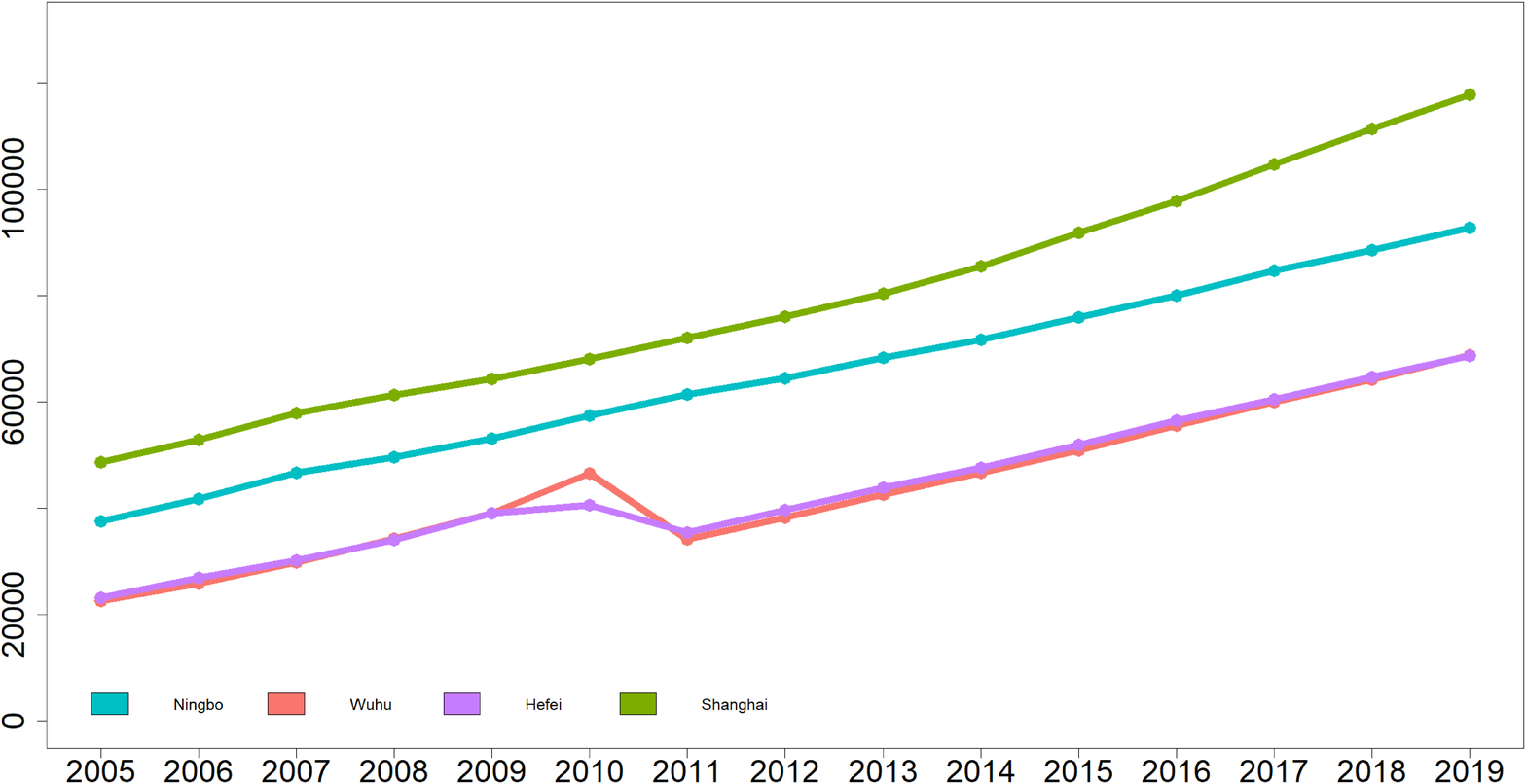}
			\caption{GDP per capita 2005 constant prices (CNY)}
		\end{subfigure}
		\begin{subfigure}[b]{0.49\textwidth}
			\centering
			\includegraphics[width=\textwidth]{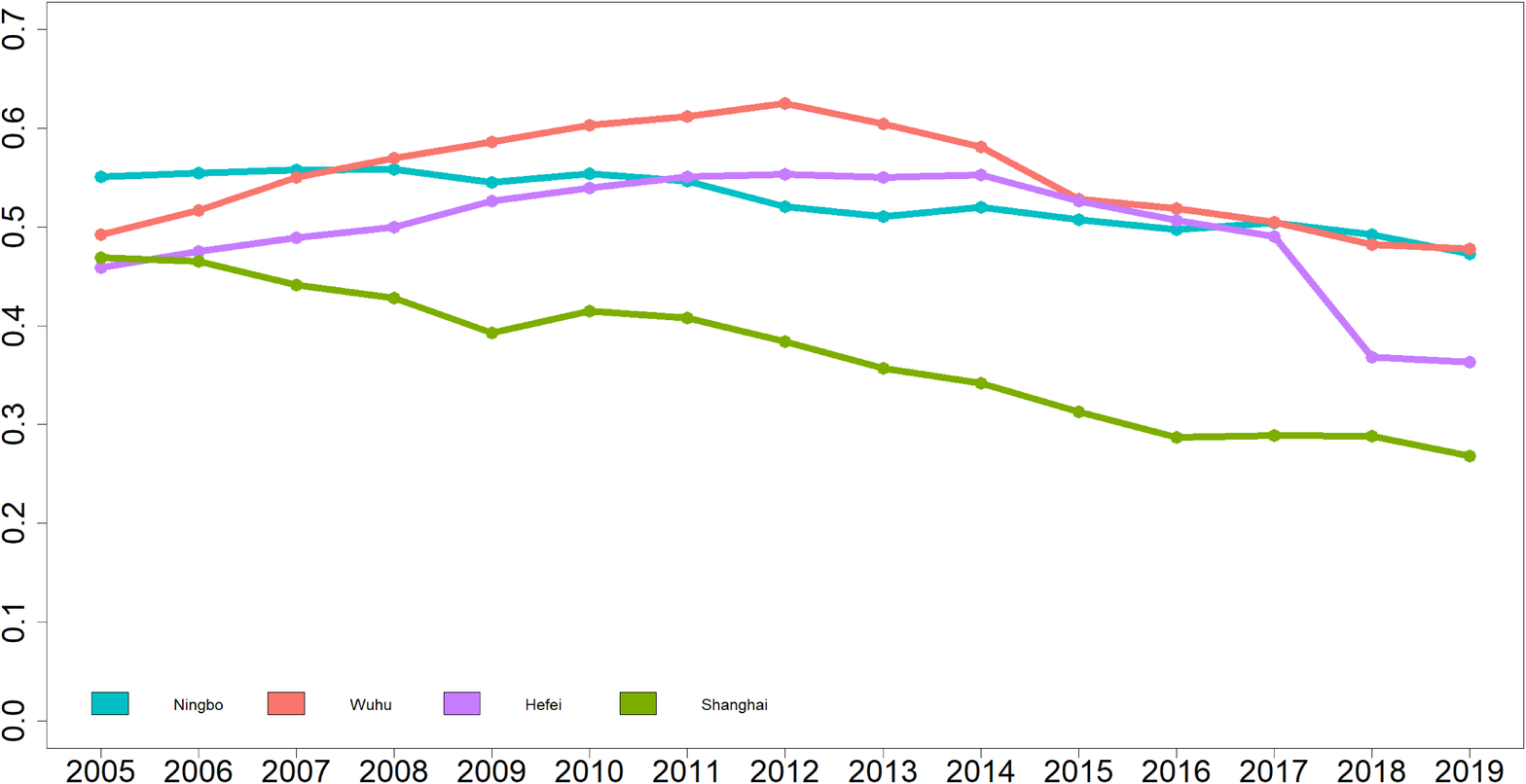}
			\caption{Percentage of secondary sector (Proportion)}
		\end{subfigure}
		\begin{subfigure}[b]{0.49\textwidth}
			\centering
			\includegraphics[width=\textwidth]{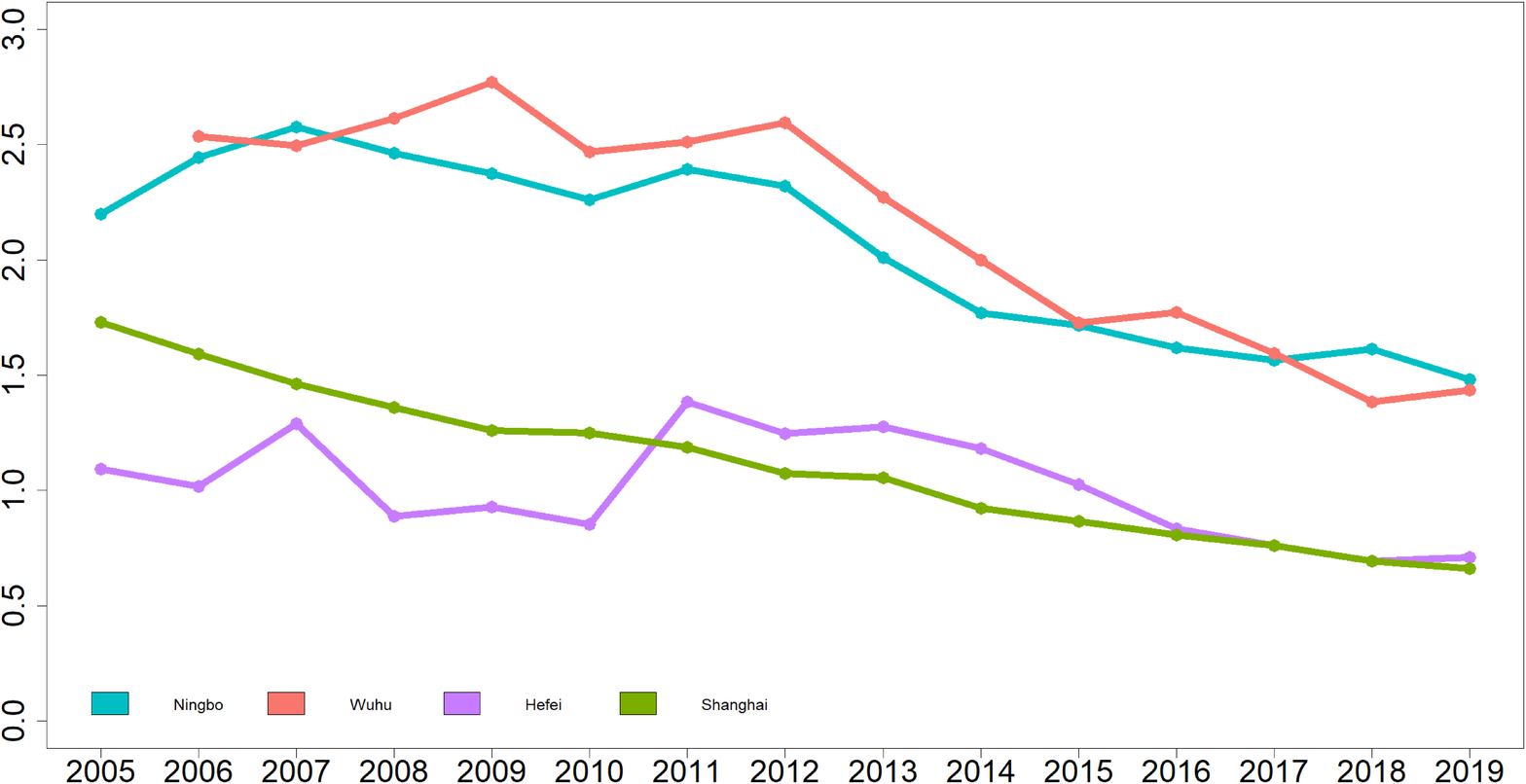}
			\caption{Carbon emission intensity  (ton/10,1000 CNY)}
		\end{subfigure}
		\caption{Green, purple, cyan and orange represent Shanghai, Hefei, Ningbo and  Wuhu respectively. }
		\label{fig:1}
	\end{figure}

	Having collected the data in  Table~\ref{tab:1}, we can calculate population ($P$),  energy intensity ($E$), GDP per capita ($A$) and percentage of secondary sector ($I$). In addition, we have calculated carbon intensity (i.e. carbon emissions $C$ divided by GDP). The results are shown in Figure~\ref{fig:1}, due to space constraints we only show data of four representative cities: Ningbo, Shanghai, Hefei and Wuhu. The figures show that:\\
	1. The energy intensity of all four cities is on a downward trend, with Wuhu decreasing faster and Shanghai having the lowest energy intensity. \\
	2. These four cities are currently experiencing slow but steady population growth, with Shanghai having the largest population and Wuhu having the smallest. \\
	3. All four cities maintained high growth rates in GDP per capita.\\
	4. Wuhu and Ningbo currently have a higher share of secondary industries, while Shanghai's share is declining.\\
	5. The carbon emission intensity of all four cities is decreasing  currently. The carbon emission intensity of Shanghai and Hefei are close to each other; Ningbo and Wuhu are close to each other.
	
	To sum up, Shanghai is China's megalopolis, with a large population and a developed economy. High energy-consuming and high-emission industries such as industry account for a relatively small proportion of the economy as a whole, and the economic focus has shifted to service-oriented industries such as finance. But in fact, most Chinese cities like Wuhu, with a medium population, are still dominated by secondary industries and their economies are in a state of high growth, which brings with it a higher carbon emission intensity. This is the reason for our subsequent focus on Wuhu.

	\section{Comparison of modeling methods: fitting and prediction accuracy}\label{sec:fitandpre}

	We used the four model-fitting methods described in Section~\ref{sec:method} to fit the data of the ten cities from 2005 to 2019 and evaluate the goodness of fit of these four methods by comparing simulated emission values with actual emission values.
	In addition, we consider whether to include city differences as an influencing factor. This is because cities will differ in their expression of the influencing factors; for example, a city using clean energy will emit less carbon than a city using fossil fuels, even for the same energy intensity.
	We treat the city as a variable added to the model when city differences are considered and do not include the city variable when differences are not considered.
	
	
	We illustrate the goodness of fit here by showing the one-to-one plot in Figure~\ref{fig:1}, where we let the vertical coordinates be the true emission values and the horizontal coordinates the fitted emission values, so that if the constituent coordinates are closer to a line with a slope of 1 from the origin, the better the fit is. Intuitively, the results of the kernel method are closer to the true values with or without accounting for city differences. Random forest and neural network also perform well when city differences are considered. 
	In addition, we also provide the mean squared error (MSE) and bias in Table~\ref{tab:2}.
	We find that the kernel method has the minimum MSE and bias both with and without considering city differences, followed by the neural network method with considering city differences.
	
	This does not indicate that the kernel method is optimal, as it may be over-fitted. In fact, what is more important is that we are more interested in the predictive power of the model. We need predictions of carbon emissions under different future scenarios to provide  policy recommendations.
	
	
	\begin{figure}[htp]
		\centering
		\begin{subfigure}[b]{0.45\textwidth}
			\centering
			\includegraphics[width=\textwidth]{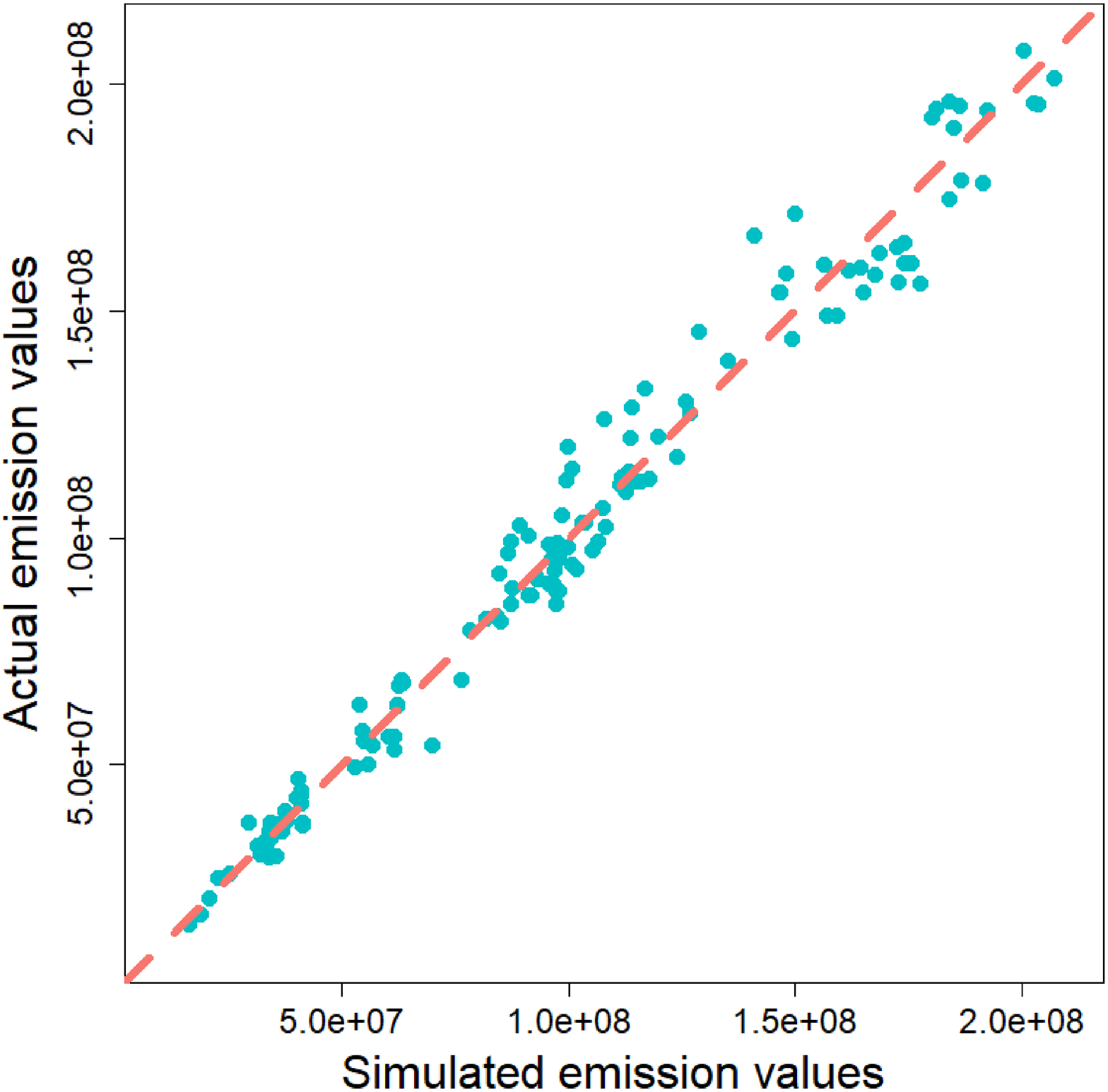}
		\end{subfigure}
		\begin{subfigure}[b]{0.45\textwidth}
			\centering
			\includegraphics[width=\textwidth]{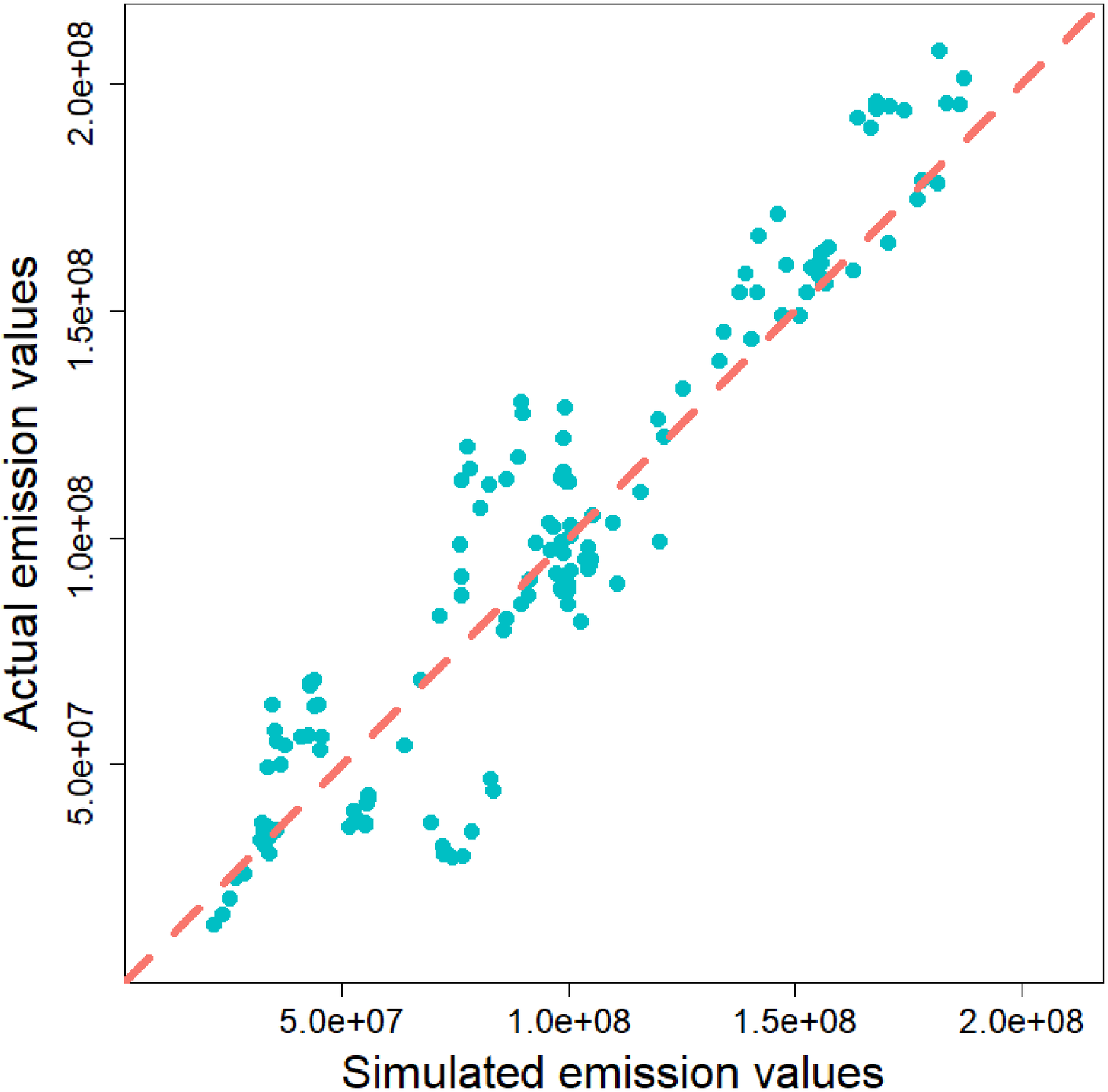}
		\end{subfigure} \\
		
		\begin{subfigure}[b]{0.45\textwidth}
			\centering
			\includegraphics[width=\textwidth]{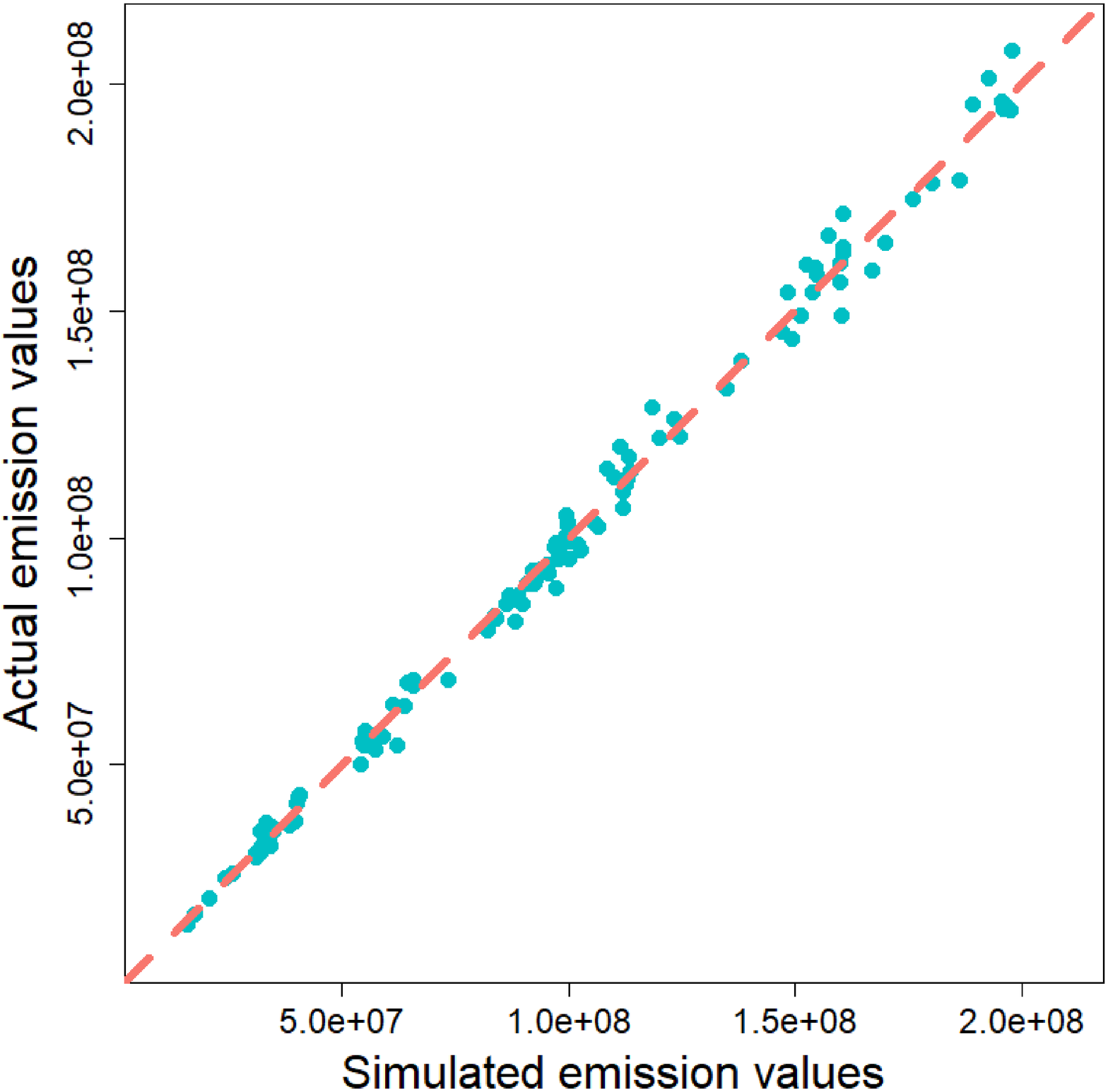}
		\end{subfigure}
		\begin{subfigure}[b]{0.45\textwidth}
			\centering
			\includegraphics[width=\textwidth]{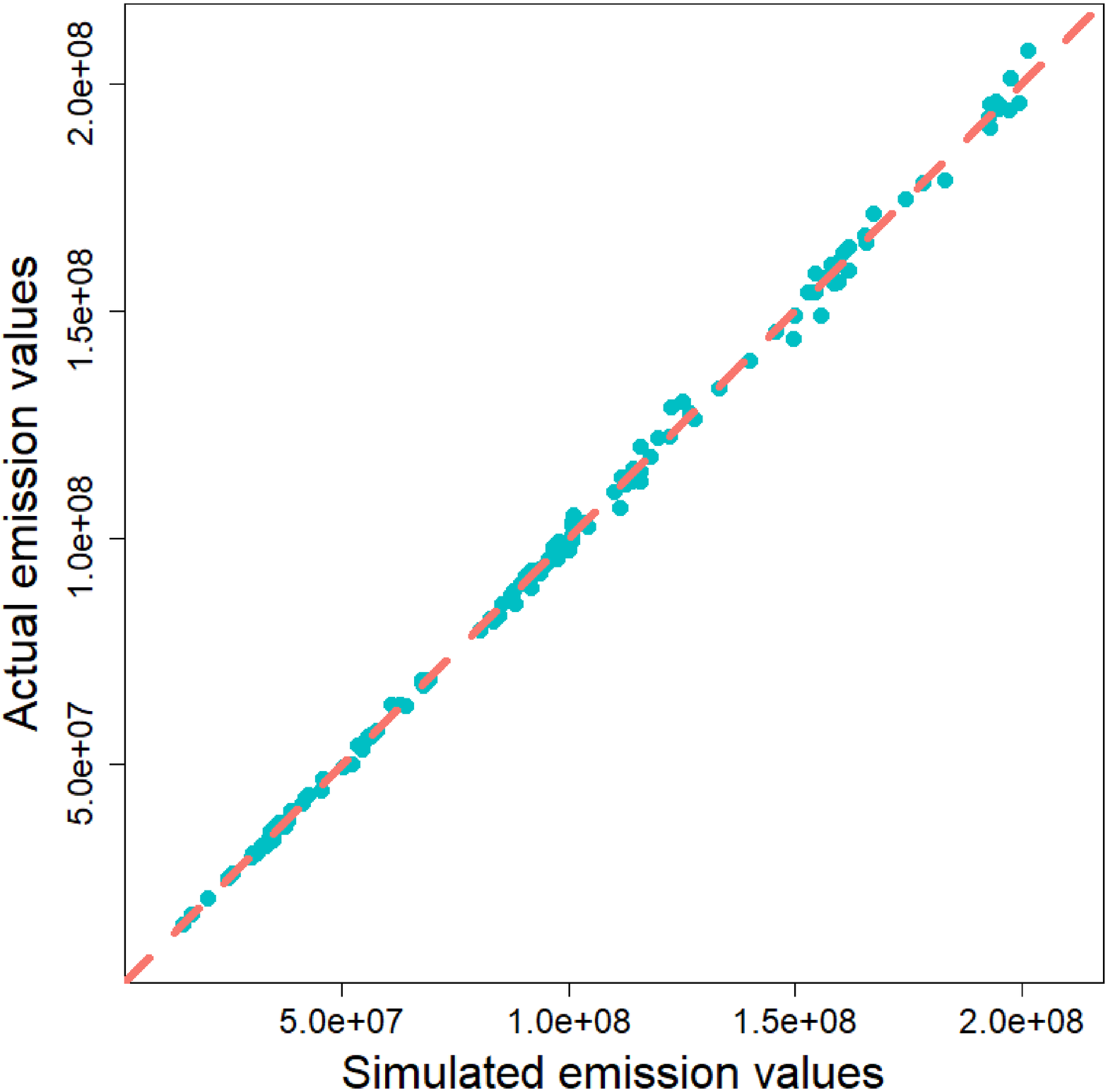}
		\end{subfigure} \\
		\begin{subfigure}[b]{0.45\textwidth}
			\centering
			\includegraphics[width=\textwidth]{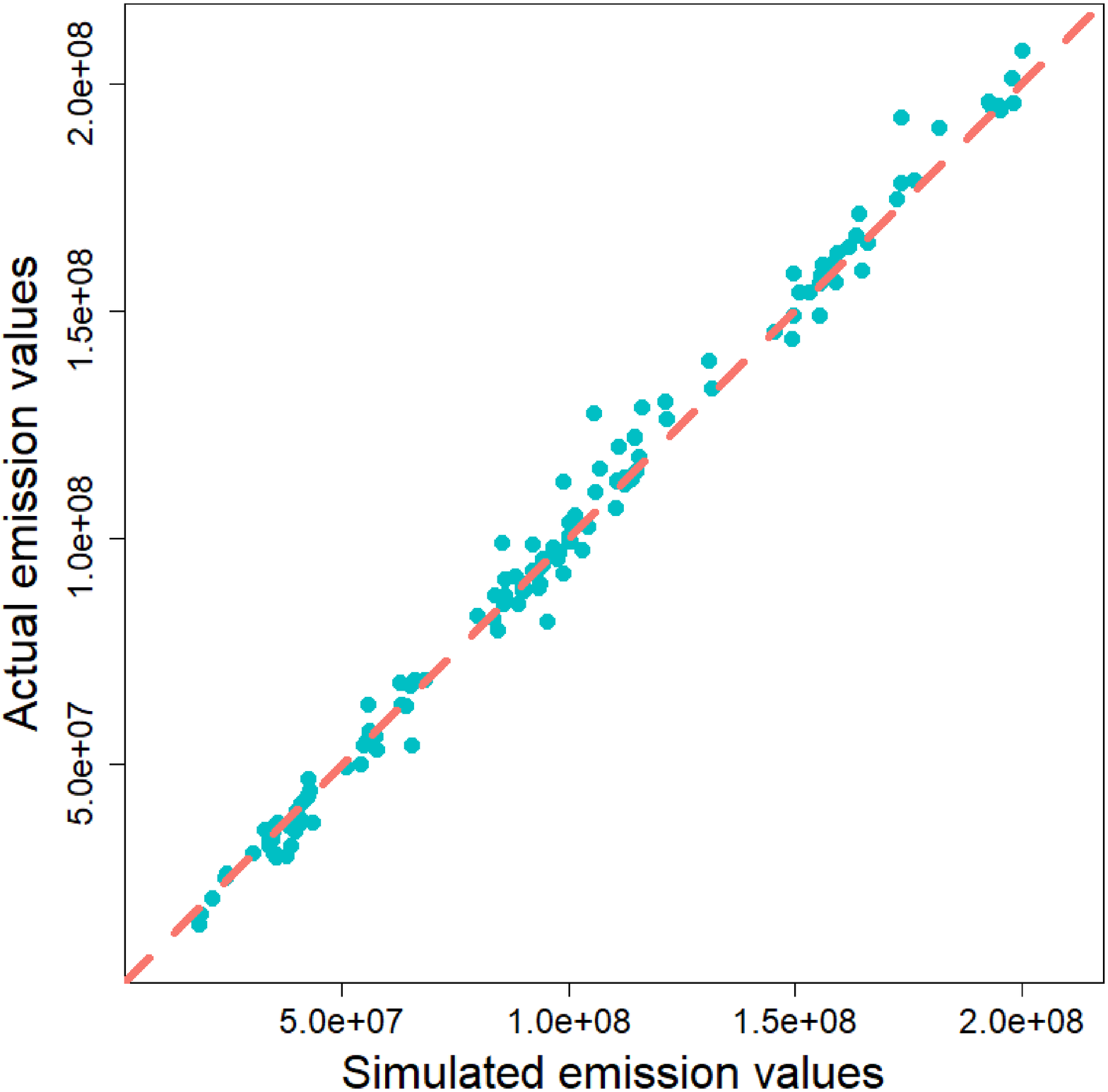}
		\end{subfigure}
		\begin{subfigure}[b]{0.45\textwidth}
			\centering
			\includegraphics[width=\textwidth]{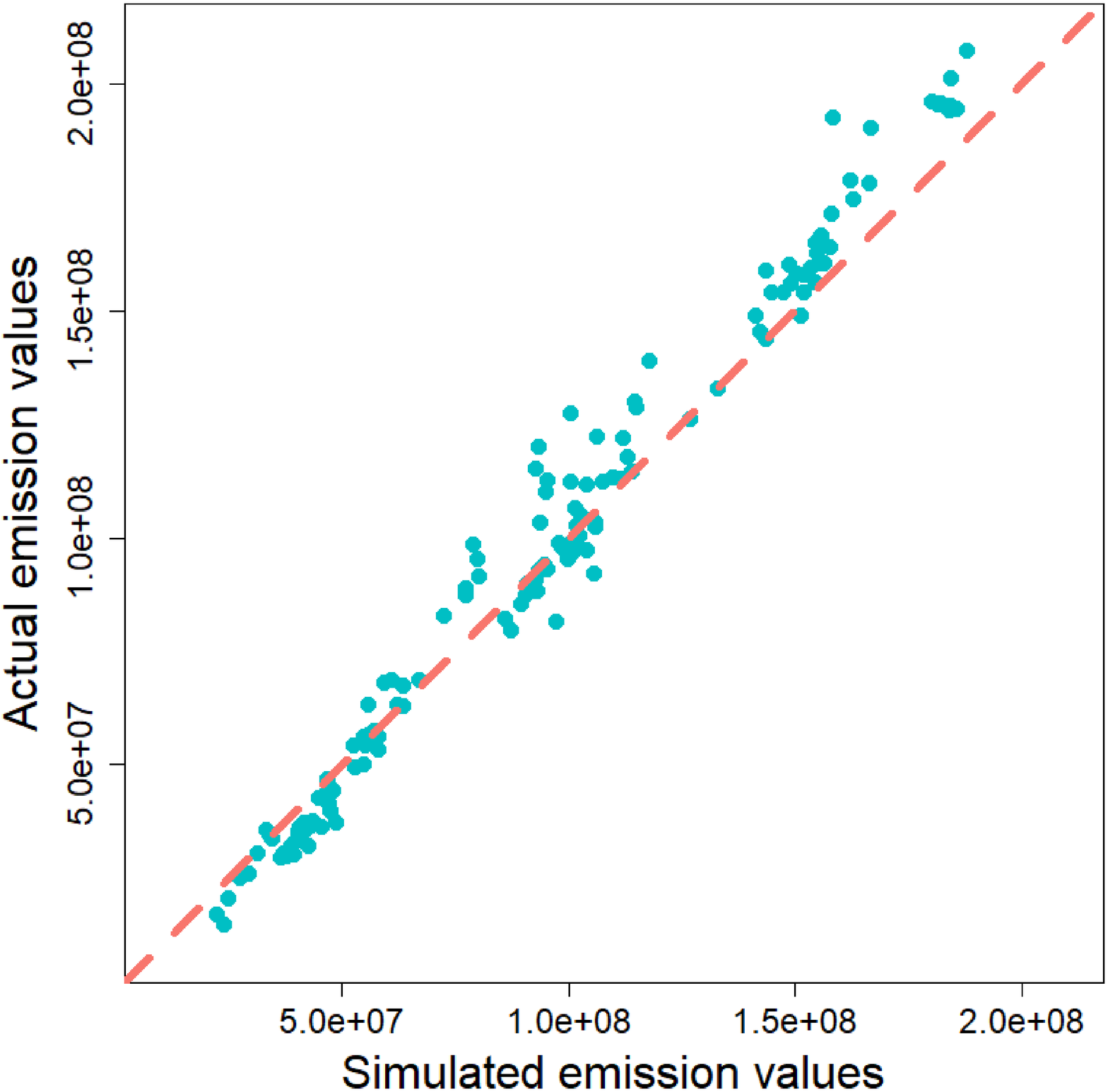}
		\end{subfigure}\\
		
		\begin{subfigure}[b]{0.45\textwidth}
			\centering
			\includegraphics[width=\textwidth]{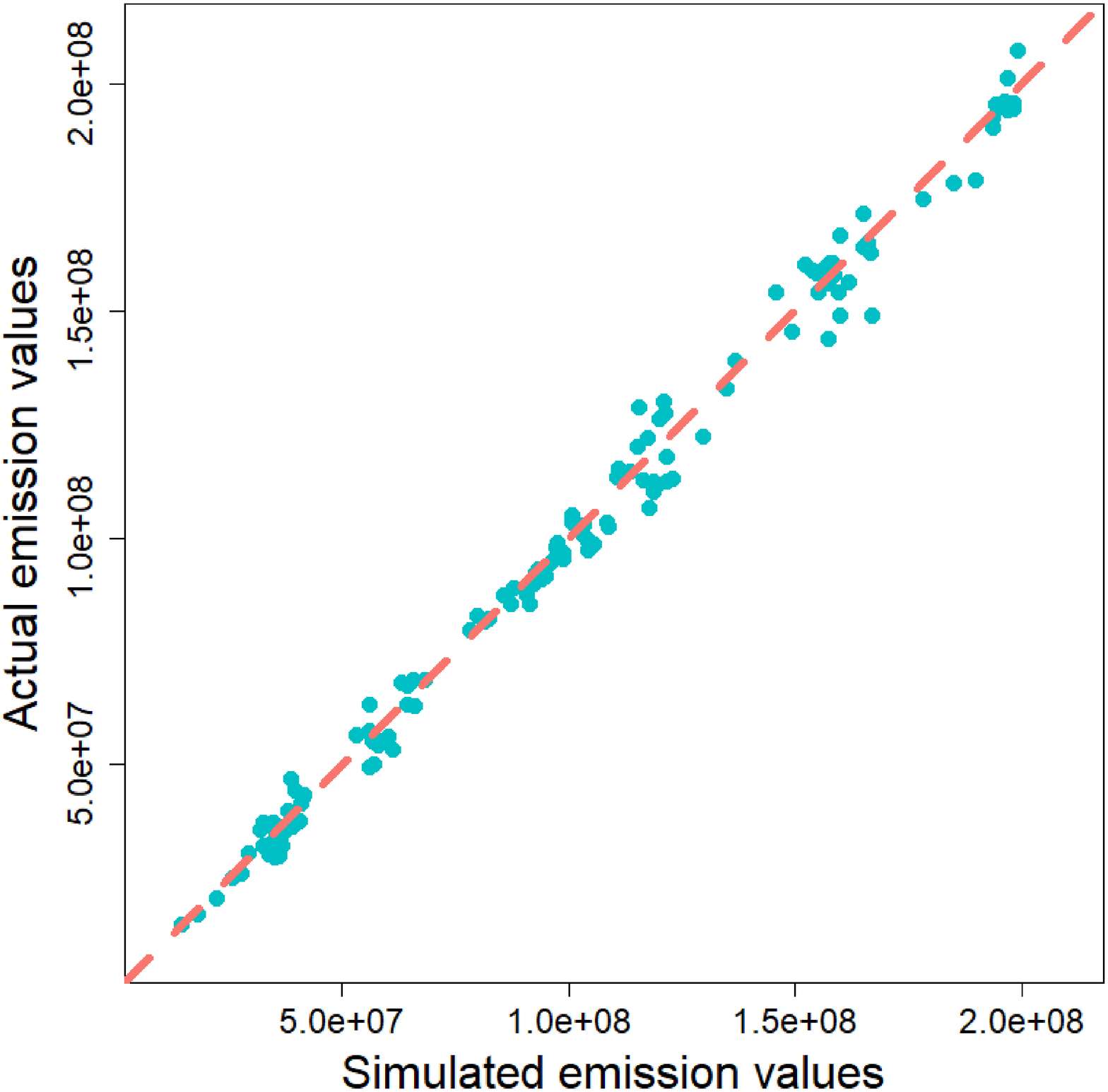}
		\end{subfigure}
		\begin{subfigure}[b]{0.45\textwidth}
			\centering
			\includegraphics[width=\textwidth]{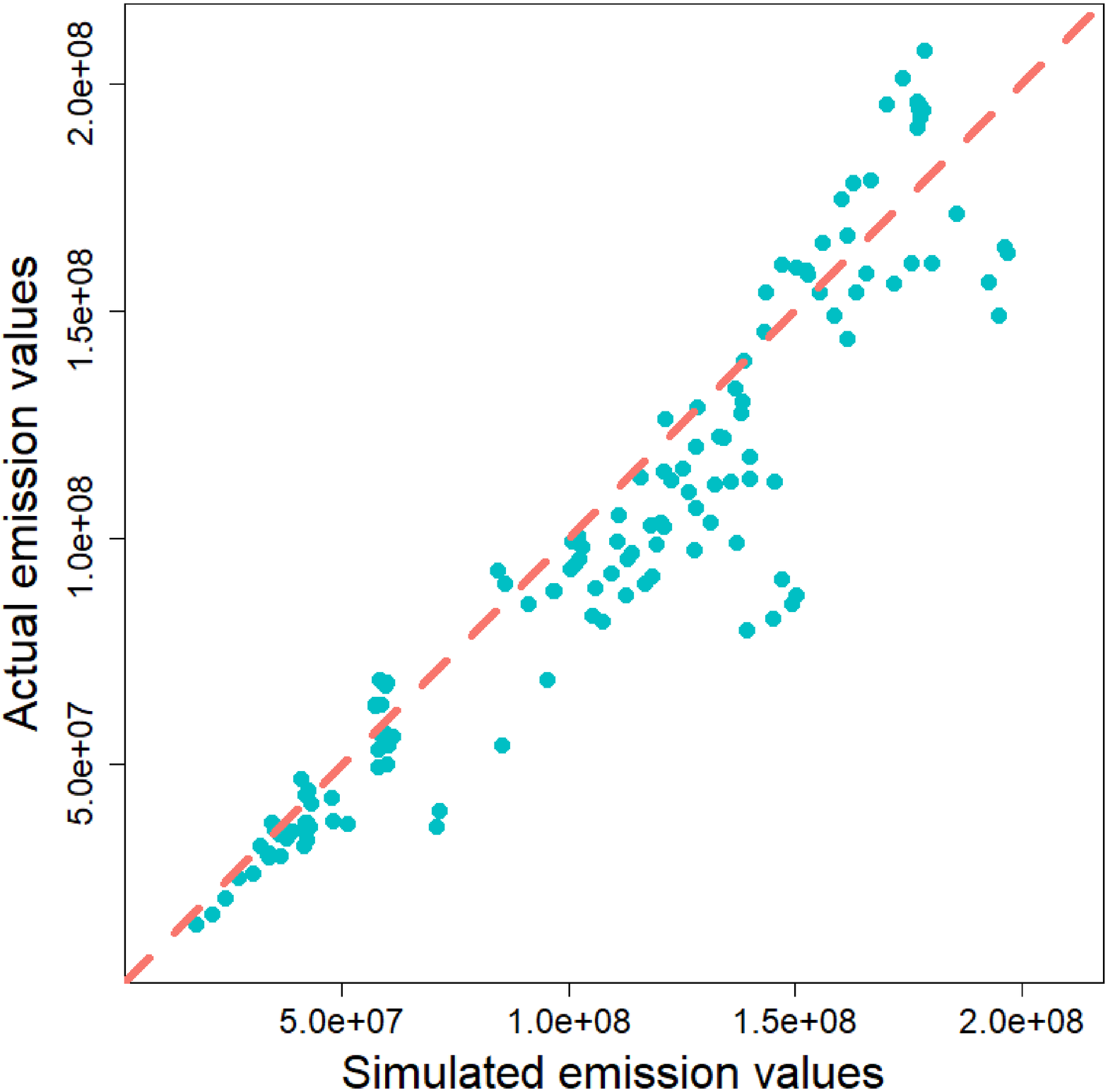}
			
		\end{subfigure}
		\caption{ One-by-one plots of fitted and actual carbon emissions for the different methods. The red dashed line is from the origin  with a slope of 1. The first, second, third  and fourth  rows  represent the results of the  linear regression, kernel regression, random forest, and neural network, respectively. the first and second columns  represent the results of considering city differences and not considering, respectively.}
		\label{fig:2}
	\end{figure}
	
	We next investigate the predictive power of the four methods. We note that when city differences are considered, the fit results are better than those without city differences, except for the kernel method, which is close to that when it is not considered. This indicates that differences between cities do exist. Therefore, in the following study, we consider city differences in all methods.
	We train the model using only data from 2005 to 2017, make predictions for 2018 and 2019, and compare the predictions with real emissions.  We show the actual emission values and the predicted emissions from the four methods in Figure~\ref{fig:2}. We found that the predicted values of neural networks were mostly close to the true values, and the results of random forest methods were often far from the true values, which happened occasionally for kernel regression and linear regression.
	In addition, bias and MSE are provided in Table~\ref{tab:2}. Neural network has the minimum MSE and bias, which is a significant advantage over other methods.
	Kernel methods that perform well in fitting tasks are far inferior to neural networks in prediction tasks. In fact, kernel regression methods suffer from curse of dimensionality in the face of multiple variables~\cite{bengio2005curse}. Their great fitting results may due to the over-fitting.
	
	Combining the results of both fitting and prediction, neural networks have better performance and we hence use them throughout the rest of  this article.
	

	\begin{figure}[htp]
		\centering
		\begin{subfigure}[b]{1\textwidth}
			\centering
			\includegraphics[width=1\textwidth]{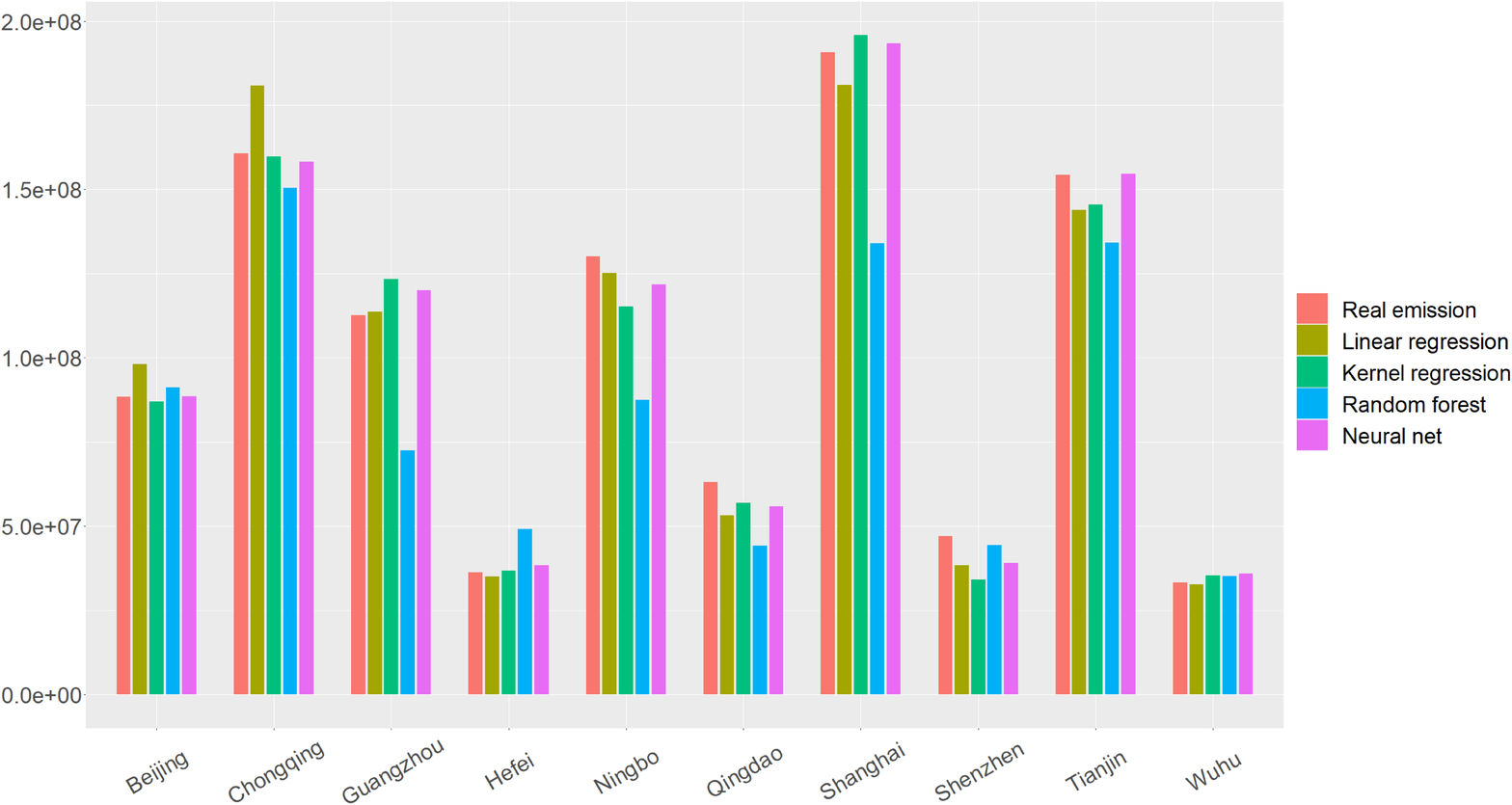}
		\end{subfigure}\\
		\hspace{0.2cm}
		\begin{subfigure}[b]{1\textwidth}
			\centering
			\includegraphics[width= 1\textwidth]{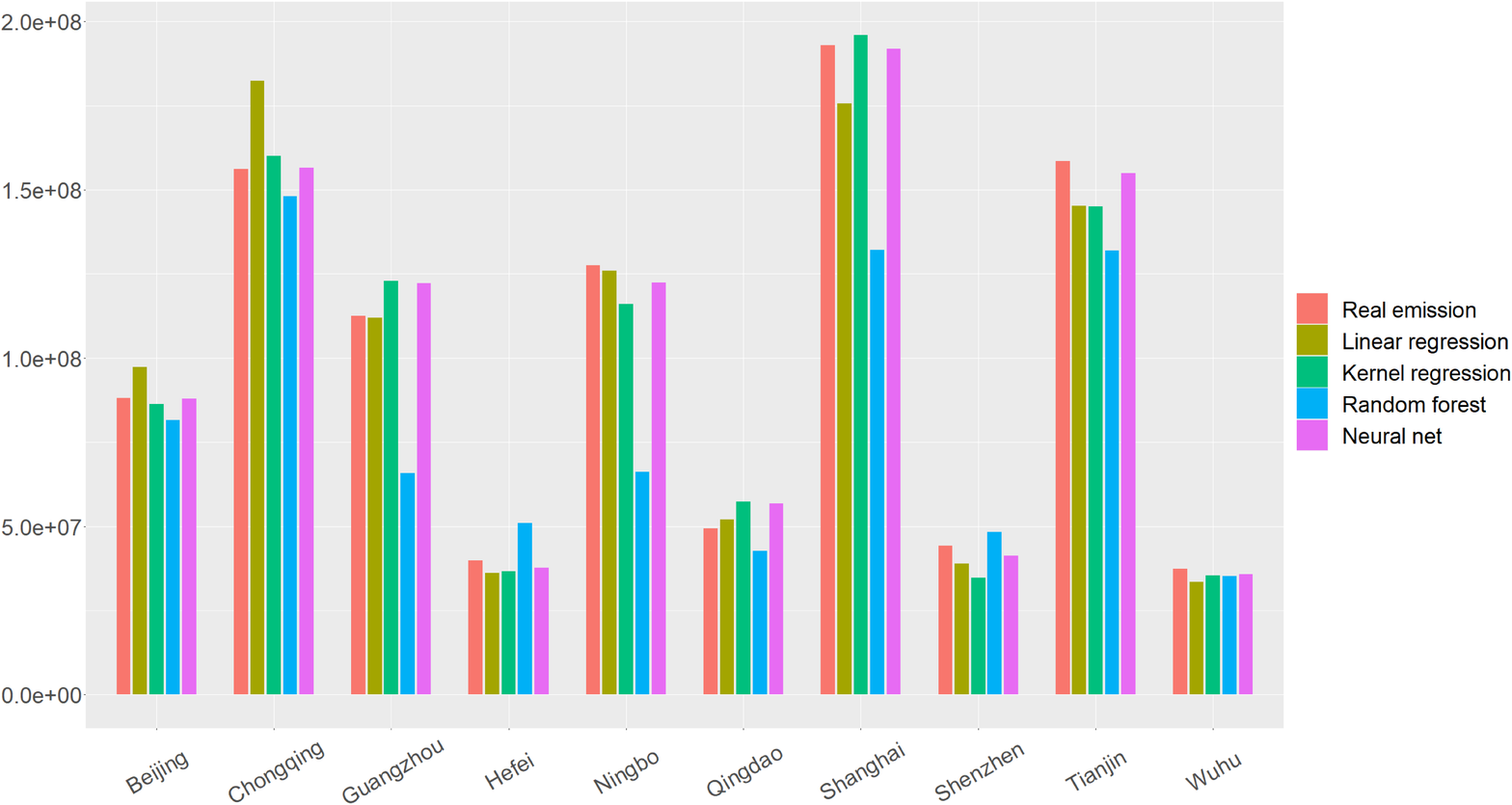}
		\end{subfigure} 
		\caption{Real and predicted carbon emission values for the four methods. Upper panel: results for the year of 2018. Lower panel: results for the year of 2019.}
		\label{fig:3}
		
	\end{figure}
	
	\begin{table}[htbp]
		\centering
		\begin{tabular}{c|c|ccc}
			\hline
			\multicolumn{1}{c}{} & \multicolumn{1}{c}{} & Method & MSE  & Bias \bigstrut\\
			\hline
			\multirow{8}[16]{*}{Fitting} & \multirow{4}[8]{*}{City differences} & Linear regression &$  6.29\times 10^{13} $ & $ 5.94\times 10^6 $ \bigstrut\\
			\cline{3-5}          &       & Kernel Regression &  $ \boldsymbol{1.53\times 10^{13}} $ & $ \boldsymbol{2.89\times 10^6} $ \bigstrut\\
			\cline{3-5}          &       & Random Forest & $ 2.60\times 10^{13 }$ & $ 3.56\times 10^6 $ \bigstrut\\
			\cline{3-5}          &       & Neural Network & $ 2.39\times 10^{13} $ & $ 3.78\times 10^6 $ \bigstrut\\		
			\cline{2-5}          & \multirow{4}[8]{*}{No city differences} & Linear regression & $ 3.48\times 10^{14} $ & $ 1.43\times 10^7  $\bigstrut\\
			\cline{3-5}          &       & Kernel Regression & $ \boldsymbol{1.46\times 10^{13}} $ & $ \boldsymbol{2.85\times 10^6} $ \bigstrut\\
			\cline{3-5}          &       & Random Forest & $ 8.40\times 10^{13 }$ & $ 6.97\times 10^6 $ \bigstrut\\
			\cline{3-5}          &       & Neural Network & $ 3.71\times 10^{14} $ & $ 1.39\times 10^7 $ \bigstrut\\
			\hline
			\multicolumn{5}{c}{} \bigstrut\\
			\hline
			\multirow{8}[16]{*}{Prediction} & \multirow{4}[8]{*}{2018} & Linear regression & $ 9.04\times 10^{13} $ & $ 7.63\times 10^6 $ \bigstrut\\ 
			\cline{3-5}          &       & Kernel Regression & $ 6.89\times 10^{13}  $& $ 6.23\times 10^6  $\bigstrut\\
			\cline{3-5}          &       & Random Forest & $ 6.56\times 10^{14 }$ & $ 1.95\times 10^7 $ \bigstrut\\
			\cline{3-5}          &       & Neural Network & $ \boldsymbol{2.64\times 10^{13}} $ & $ \boldsymbol{4.13\times 10^6} $ \bigstrut\\
			
			\cline{2-5}          & \multirow{4}[8]{*}{2019} & Linear regression &  $ 1.31\times 10^{14 }$ & $ 8.35\times 10^6 $ \bigstrut\\
			\cline{3-5}          &       & Kernel Regression & $ 6.09\times 10^{13} $ & $ 6.36\times 10^6 $ \bigstrut\\
			\cline{3-5}          &       & Random Forest & $ 1.08\times 10^{15} $  & $ 2.36\times 10^7 $ \bigstrut\\
			\cline{3-5}          &       & Neural Network & $ \boldsymbol{2.06\times 10^{13}} $  & $ \boldsymbol{3.39\times 10^6} $ \bigstrut\\
			
			\hline
		\end{tabular}%
		\caption{This table provides the predicted or fitted MSE and bias for the different methods. The smallest bias and MSE are shown in bold.}
	\label{tab:2}%
\end{table}%

\section{Neural network-based analysis and forecasting}\label{sec:wuhu}

In this section, we analyze the city of Wuhu\footnote{A city that is of research interest to our funding institute.}   as an example, focusing on the factors affecting carbon emissions in Wuhu  and making predictions. Currently, Wuhu is still an industrial city, with the secondary industry consuming close to 50\% of the energy, and over 60\% of the energy supply for the whole city is coal. This results in Wuhu having high total carbon emissions and carbon emission intensity. However, there is also some potential for emission reduction, such as improving energy efficiency, optimizing industrial structure and developing clean energy. 
Most cities in China share these similarities to Wuhu.
We further specify the analysis of the factors influencing carbon emissions in Wuhu and the  prediction of carbon emissions. This could be a guide for many Chinese cities to reduce their carbon emissions and help China as a whole to reach its carbon peak and carbon neutral targets sooner. We begin with an impact factor analysis.

In previous studies, the modeling has generally taken  linear regression approach and used the regression coefficients of the variables to determine the impact of the variables on carbon emissions. However, this is not reasonable because the difference in cities and time, coupled with the existence of interactions between variables, makes it very restrictive to use only one coefficient to account for the effect of the variables on emissions. 

Because neural network has no explicit expressions,  here we have taken a new approach to quantify the impact of different variables on emissions. That is, for a given city at a given time, all variables are entered into the model unchanged to obtain a baseline value; then when analyzing the impact of a variable on emissions, the other variables are fixed and the value of this variable is increased by 1\%, and the output of the modet al this point is compared with the baseline value for that city at that moment in time. 

The Wuhu Statistical Yearbook has been updated to 2022 to include data for 2021. We use 2021 data for Wuhu as the baseline data, and both the manipulation of the data and the forecast results are shown in Table~\ref{tab:3}.

\begin{table}[htbp]
	\centering
	\resizebox{.9\columnwidth}{!}{
		\begin{tabular}{cccccc}
			\hline
			Situations & $ P  $    & $ A  $ (CNY) & $ I   $   & $ E  $ (ton/10,000 CNY) & Forecasting emission (ton) \bigstrut\\
			\hline
			Baseline & 3,818,000 & 75,001 & 0.451 & 0.4463 & 36,659,741 \bigstrut\\
			\hline
			$ P \times 101\%  $   & 3,856,180 & 75,001 & 0.451 & 0.4463 & 36,654,857 \bigstrut\\
			\hline
			$ A   \times 101\%  $   & 3,818,000 & 75,751 & 0.451 & 0.4463 & 37,029,755 \bigstrut\\
			\hline
			$ I \times 101\%   $   & 3,818,000 & 75,001 & 0.456 & 0.4463 & 36,506,851 \bigstrut\\
			\hline
			$ E \times 101\%   $   & 3,818,000 & 75,001 & 0.451 & 0.4507 & 37,129,085 \bigstrut\\
			\hline
		\end{tabular}%
	}
	\caption{Results of the significance analysis of the impact factors. In each situation, the value of a variable is increased by $1\%$.}
	\label{tab:3}%
\end{table}%

We find that Wuhu's carbon emissions are insensitive to population, with little or no change. Higher levels of affluence lead to more emissions, which again corresponds to the results of  Dietz el at.\cite{dietz1997effects}. An increase in energy intensity also significantly increases carbon emissions. We note, however, that holding other things constant, increasing the share of secondary industry decreases carbon emissions, which we believe is because a higher share of secondary industry does not raise energy consumption, implying that industry is transferred towards less energy consumption, which generally represents less emissions. 

\begin{figure}[htp]
	\centering
	\includegraphics[width=0.97\textwidth]{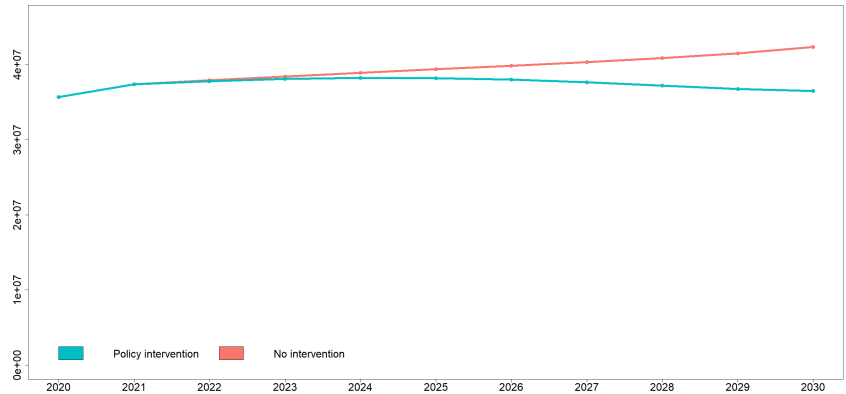}
	\caption{This is the Wuhu carbon emission prediction curve (to 2030).}
	\label{fig:4}
\end{figure}

In the following, we make predictions for Wuhu's carbon emissions. First, we give a baseline scenario based on Wuhu's historical data and the government's current targets. We do not specify 2020 and 2021 any further because they are available.
For 2022 to 2023 we assume a 2\% year-on-year decrease in energy intensity, a 5\% year-on-year increase in GDP per capita, a constant population and a 1\% year-on-year decrease in the share of secondary industry. It should be explained that the population remains unchanged due to previous Chinese family planning policies and the siphoning effect of large cities on Wuhu, which we assume will remain largely unchanged. In addition, we assume that policy interventions are made to promote industrial upgrading and technological upgrading at the expense of some economic growth, so we assume a 4\% year-on-year decrease in energy intensity, a 3\% year-on-year increase in GDP per capita and a 2\% year-on-year decrease in the share of secondary industry and  that population remains constant. We put the results in Figure~\ref{fig:4}. We find that Wuhu's carbon emissions will continue to grow without policy intervention, while with policy intervention, they peak around 2025 and then trend downwards.

Based on the research findings, several policy implications can be drawn. Given the significance of industrial structure on carbon emissions, great measures should be taken to optimize the existing industrial structure. On the one hand, we should limit the development of high energy-consuming and high-polluting industries, vigorously develop green and clean industries and high-tech enterprises, so as to reduce the proportion of polluting enterprises in the secondary industry, increase the proportion of clean and high-tech industries in the secondary industry, and improve the greenness of the secondary industry. On the other hand, we should vigorously develop the tertiary industry, increase the proportion of the service industry in the overall economic structure, and moderately reduce the proportion of the secondary industry in the economic structure. Furthermore, energy intensity and carbon emissions are closely related, and measures should be taken to improve energy efficiency and reduce energy intensity. For example, enterprises should be encouraged to strengthen the research, development and application of energy-saving technologies to improve their energy efficiency and resource utilization efficiency. Meanwhile, the application of emerging digital technologies such as big data, cloud computing and artificial intelligence needs to be accelerated to further optimize production processes, increase energy-saving potential and reduce carbon intensity.

\section{Conclusion, implication and limitation}\label{sec:conclusion}


Inspired by STIRPAT, we consider not only linear regression but also the use of statistical and machine learning non-parametric methods to model carbon emissions.
Specifically, we model the carbon emission regression using three nonparametric methods: kernel regression, random forest, and neural network, using the factors proposed in STIRPAT  as independent variables and carbon emissions as dependent variable, and finally find that the neural network has the best performance (based on data for ten cities from 2005 to 2019). In addition, we quantitatively analyze the factors influencing carbon emissions based on this model and forecast future emissions, which leads to policy recommendations.  In general, this study provides new ideas and approaches for carbon emission analysis.

Though this research is meaningful, several aspects should be noted. First, at the time of modeling, there are various factors affecting carbon emissions, while only four influencing factors are considered in this research. Other influencing factors such as energy structure, regional investment scale, and foreign direct investment are not considered. The follow-up study can consider these factors.
Second, due to limited access to the carbon emissions data at the city level, the sample used for the model validation  is relatively small, which affects the generality of the results to some extent. In the following research, the research sample needs to be further expanded to further enhance the generalizability of the findings.  



\bibliographystyle{gbt7714-numerical}
\bibliography{carbonPaper}


\newpage

\end{document}